\documentstyle[amsmath,11pt]{article}

\date{\empty}

\begin{document}

\title{\bf Primordial magnetic seed field
amplification by gravitational waves: comment on gr-qc/0503006}

\author{Christos G. Tsagas\\ {\small Section of Astrophysics,
Astronomy and Mechanics, Department of Physics}\\ {\small Aristotle
University of Thessaloniki, Thessaloniki 54124, Greece}}

\maketitle

\begin{abstract}
We consider the amplification of cosmological magnetic fields by
gravitational waves as it was recently presented in~\cite{BZDM}.
That study confined to infinitely conductive environments, arguing
that on spatially flat Friedmann backgrounds the gravito-magnetic
interaction proceeds always as if the universe were a perfect
conductor. We explain why this claim is not correct and then
re-examine the Maxwell-Weyl coupling at the limit of ideal
magnetohydrodynamics. We find that the scales of the main results
of~\cite{BZDM} were not properly assessed and that the incorrect
scale assessment has compromised both the physical and the numerical
results of the paper. This comment aims to clarify these issues on
the one hand, while on the other it takes a closer look at the
gauge-invariance and the nonlinearity of~\cite{BZDM}.
\end{abstract}

\section{Introduction}
The interaction between electromagnetic fields and gravitational
waves and the possible energy transfer between the Weyl and the
Maxwell fields have a long research history. A mechanism for the
amplification of large-scale magnetic fields by gravity waves of
similar size soon after inflation was recently proposed
in~\cite{TDM,T}. In the poorly conductive environment of early
reheating the analysis indicated a resonant magnetic amplification
proportional to the square of the field's scale (see~\cite{T} for
details). This meant that Weyl-curvature distortions could provide a
very efficient early-universe dynamo of superhorizon-sized magnetic
fields. For example, fields with a current comoving scale of
approximately 10~kpc and a strength of $\sim10^{-34}$~G, like those
produced in~\cite{D}, could be amplified by many orders of magnitude
by the end of reheating.

The same gravito-magnetic interaction has been applied to infinitely
conductive cosmologies in~\cite{BZDM}. Central to that study is the
claim that on spatially flat Friedmann-Robertson-Walker (FRW)
backgrounds the Maxwell-Weyl coupling proceeds always as if the
universe were a perfect conductor. We argue that this is not the
case and explain why the above mentioned paper arrived at the
opposite result. In addition, we find serious problems in the scale
assessments of~\cite{BZDM}. These have prevented the authors from
recognising the main effect of the interaction, namely the
large-scale superadiabatic amplification of the $B$-field, and
compromised their results. With the present comment we draw
attention to these issues by correcting the mathematics, where
necessary, and by clarifying the physics. We also re-examine and
question the gauge-invariance and the nonlinearity of the formalism
proposed in~\cite{BZDM}.

\section{On the electric curl and the role of conductivity}
Section III.B of~\cite{BZDM} argues that the Maxwell-Weyl coupling
proceeds unaffected by the conductivity of the medium and as if the
universe were a perfect conductor. This follows the claim that the
gravitationally induced, second order, electric field ($E_a$) is
curl-free if the zero order FRW model has flat spatial sections.
This is not the case, however, because
\begin{equation}
({\rm curl}E_a)^{\cdot}=-\Theta{\rm curl}E_a+ {\cal
R}_{ab}\tilde{B}^b- {\rm curl}{\cal J}_a- {\rm D}^2B_a\,,
\label{eq:curlEdot}
\end{equation}
at second order. Here $\Theta$ is the volume expansion, ${\cal
R}_{ab}$ is the first-order 3-Ricci tensor, $\tilde{B}_a$ is the
first order magnetic field, ${\cal J}_a$ is the spatial current and
${\rm D}^2={\rm D}_a{\rm D}^a$ is the 3-dimensional Laplacian
operator. Therefore, even when ${\rm curl}E_a$ initially vanishes,
there are sources in the right-hand side of~(\ref{eq:curlEdot}) that
will generally lead to a nonzero electric curl. Switching the latter
off is not a consistent constraint. Note that the current term
depends on the conductivity of the medium, while the 3-curvature
distortions and the magnetic field fluctuations are caused by the
presence of gravity-wave perturbations. These set the field lines
into motion causing magnetic fluctuations that produce an electric
component. The latter has ${\rm curl}E_a\neq0$ because of
(\ref{eq:curlEdot}). This means that the gravito-magnetic
interaction does not always proceed as if the conductivity of the
universe were infinite. In~\cite{BZDM}, the authors arrived at the
opposite conclusion because they considered the second
time-derivative of ${\rm curl}E_a$ instead of the first, namely (see
Eq.~(29) in~\cite{BZDM})
\begin{equation}
({\rm curl}E_a)^{\cdot\cdot}+{\textstyle{7\over3}}\Theta({\rm
curl}E_a)^{\cdot}- {\rm D}^2{\rm curl}E_a+
\left[{\textstyle{7\over9}}\Theta^2+{\textstyle{1\over6}}(\mu-9p)
+{\textstyle{5\over3}}\Lambda\right]{\rm curl}E_a={\rm curl}K_a\,.
\label{eq:D1}
\end{equation}
In the above, which is said to hold at the second perturbative
level, the pair $\mu$ and $p$ represents the density and the
pressure of the matter, $\Lambda$ is the cosmological constant and
$K_a$ is a gravito-magnetic source term. Arguing that ${\rm
curl}K_a$ vanishes when the FRW background is spatially flat (see
Eq.~(30) in~\cite{BZDM}), the paper claims that the electric field
will remain curl-free if it was so initially. However, this a priori
sets $({\rm curl}E_a)^{\cdot}=0$ in Eq.~(\ref{eq:D1}), although the
latter is not necessarily the case because of expression
(\ref{eq:curlEdot}).

By switching the electric curl off the authors have inadvertently
confined their study to idealised perfectly conductive universes,
bypassing all plasma effects and the implications of finite
conductivity. Because of that~\cite{BZDM} cannot follow the
gravito-magnetic interaction in the poorly conductive stages of
early reheating and therefore the comparison with~\cite{TDM,T} was
inappropriate.

\section{On the scale assessment and the nature of the
amplification}
Restricting to perfectly conducting cosmological environments we
have (see \S~II.C, IV.B in~\cite{BZDM})
\begin{equation}
\dot{B}_a+ {\textstyle{2\over3}}\Theta B_a= \sigma_{ab}\tilde{B}^b
\equiv I_a\,,  \label{eq:dotB}
\end{equation}
where $B_a$ is the total magnetic field, $\tilde{B}_a$ is the
original and $\sigma_{ab}$ is the transverse shear component. Then,
the gravitationally induced $B$-fields during the radiation era is
\begin{equation}
B_{R}=\tilde{B}_0\left(\frac{a_0}{a}\right)^2
\left\{1+{\textstyle{2\over3}}\left(\frac{\sigma}{H}\right)_0
\left[\left(\frac{a_0}{a}\right)-1\right]
+{\textstyle{5\over6}}\left(\frac{\sigma}{H}\right)_0
\left[\left(\frac{a}{a_0}\right)^2-1\right]\right\}\,,
\label{eq:D2}
\end{equation}
while for dust and late reheating we have
\begin{equation}
B_{D/RH}=\tilde{B}_0\left(\frac{a_0}{a}\right)^2
\left\{1+{\textstyle{2\over3}}\left(\frac{\sigma}{H}\right)_0
\left[\left(\frac{a_0}{a}\right)^{3/2}-1\right]
+2\left(\frac{\sigma}{H}\right)_0
\left[\left(\frac{a}{a_0}\right)-1\right]\right\}\,,  \label{eq:D3}
\end{equation}
(see Eqs.~(46), (47) in~\cite{BZDM}). Above $a$ is the scale factor,
$H$ is the Hubble parameter and the zero suffix marks the onset of
the gravito-magnetic interaction. These expressions were treated as
infinite wavelength solutions (see \S~IV.B.1 in~\cite{BZDM}), when
in reality they cover all finite superhorizon
scales.\footnote{Assuming dust and setting $\tau=t/t_0$ the
gravito-magnetic source term in the right-hand side of (\ref{eq:D2})
reads
\begin{equation}
I_{(\ell)}(\tau)=
\left[D_1J_{5/2}\left(2\ell\tau^{1/3}/a_0H_0\right)
+D_2Y_{5/2}\left(2\ell\tau^{1/3}/a_0H_0\right)\right]\tau^{-5/2}\,,
\label{eq:D4}
\end{equation}
where $\ell$ the wavenumber of the gravitational wave (see Eq.~(48)
in~\cite{BZDM}). Finite superhorizon scales have
$\ell\tau^{1/3}/a_0H_0\ll1$ and the series expansion of the Bessel
functions (with the initial conditions of~\cite{BZDM}) give
\begin{equation} I(\tau)=
2\sigma_0\tilde{B}_0\tau^{-5/3}- \sigma_0\tilde{B}_0\tau^{-10/3}\,.
\label{eq:D5}
\end{equation}
Inserted into Eq.~(\ref{eq:D2}) the above leads to solution
(\ref{eq:D3}), which therefore applies to all finite super-Hubble
lengths and not only to infinite wavelengths. Similarly one can show
that (\ref{eq:D3}) also covers all finite superhorizon lengths.}
This incorrect scale assessment has led to the sequence of problems
that we will discuss below. Here we note that the domains of
(\ref{eq:D2}) and (\ref{eq:D3}) are essential for the additional
reason that both results show a deviation from the standard $a^{-2}$
magnetic decay-law and a superadiabatic amplification of the field
(i.e.~$B=$~constant for radiation and $B\propto a^{-1}$ for
dust/reheating).

Returning to~\cite{BZDM} we find (see Eq.~(51) there) that the final
solution for the gravitationally induced magnetic field, during both
the radiation and the dust eras, is
\begin{equation}
\frac{B}{\rho_{\gamma}^{1/2}}\simeq \left[1+ \frac{1}{10}
\left(\frac{\lambda_{\tilde{B}}}{\lambda_H}\right)^2_0
\left(\frac{\sigma}{H}\right)_0\right]
\left(\frac{\tilde{B}}{\rho_{\gamma}^{1/2}}\right)_0 \Rightarrow
B\simeq\tilde{B}_0\left[1+ \frac{1}{10}
\left(\frac{\lambda_{\tilde{B}}}{\lambda_H}\right)^2_0
\left(\frac{\sigma}{H}\right)_0\right]
\left(\frac{a_0}{a}\right)^2\,,  \label{eq:D6}
\end{equation}
since $\rho_{\gamma}\propto a^{-4}$ is the radiation
density.~\footnote{Expression (\ref{eq:D6}) was obtained after
replacing $\lambda_{GW}$ with $\lambda_{\tilde{B}}$ in Eqs.~(49) and
(50) of~\cite{BZDM}. However, this substitution was arbitrary
because no relation between the two scales was previously
established and the original magnetic field was given a zero
wavenumber (see \S~IV.B in~\cite{BZDM}). The latter means that
$\lambda_{\tilde{B}}$ is ill defined
(i.e.~$\lambda_{\tilde{B}}\rightarrow\infty$) and misleadingly
suggests an infinitely strong induced $B$-field in (\ref{eq:D6}).
Note that assigning a nonzero wavenumber to $\tilde{B}$ provides a
useful relation between $\lambda_B$, $\lambda_{GW}$ and
$\lambda_{\tilde{B}}$ (see Eq.~(10) in~\cite{T}).} It is also argued
(see \S~IV.B.2 after Eq.~(51) in~\cite{BZDM}) that the above spans
all finite wavelengths irrespective of their size relative to the
horizon. This cannot be correct because (\ref{eq:D6}) means
$B\propto a^{-2}$ on all scales, while (\ref{eq:D2}) and
(\ref{eq:D3}) give $B=$~constant and $B\propto a^{-1}$ on
super-Hubble lengths. Consider, for example, a large-scale $B$-mode
propagating through the radiation era and ignore any effects from
reheating. Then, Eq.~(\ref{eq:D2}) gives
\begin{equation}
B\simeq\tilde{B}_0 \left[1+\left(\frac{a}{a_0}\right)_0^2
\left(\frac{\sigma}{H}\right)_0\right] \left(\frac{a_0}{a}\right)^2=
\tilde{B}_0
\left[1+\left(\frac{\lambda_{\tilde{B}}}{\lambda_H}\right)_0^2
\left(\frac{\sigma}{H}\right)_0
\left(\frac{\lambda_H}{\lambda_{\tilde{B}}}\right)^2\right]
\left(\frac{a_0}{a}\right)^2\,.  \label{eq:saB}
\end{equation}
Well outside the horizon $\lambda_H/\lambda_{\tilde{B}}\ll1$, which
means that using (\ref{eq:D6}) instead of (\ref{eq:saB}) on those
scales will grossly overestimate the amplification of the field. The
two expressions agree only at horizon crossing when
$\lambda_{\tilde{B}}=\lambda_H$. Therefore, at best,
Eq.~(\ref{eq:D6}) works for scales that have crossed the Hubble
radius. However, the magnetic growth takes place earlier outside the
horizon and results from the superadiabatic amplification of the
field seen in (\ref{eq:D2}) and (\ref{eq:D3}). In~\cite{BZDM} the
reader is unaware of the nature of the magnetic growth and where it
occurs. The root of the problem was assigning (\ref{eq:D2}) and
(\ref{eq:D3}) to infinite wavelengths. As a result, these solutions
were sidestepped (see in particular \S~V after Eq.~(60)
in~\cite{BZDM}), the suparadiabatic nature of the amplification was
not recognised, the domain of (\ref{eq:D6}) was incorrectly assessed
and ultimately the physics of the gravito-magnetic interaction was
misrepresented.

An additional complication is that Eq.~(\ref{eq:D6}) alone cannot
provide the total amplification of a mode that went through
different epochs (e.g.~reheating and radiation), because each domain
has different growth rates (compare (\ref{eq:D2}) and
(\ref{eq:D3})). Consider an inflationary $B$-mode that crosses the
horizon in the radiation era (like that in \S~V of~\cite{BZDM}).
Assuming for simplicity that the growth is always strong, the
residual field according to (\ref{eq:D6}) is
$B=\tilde{B}_0(\sigma/H)_0(\lambda_{\tilde{B}}
/\lambda_H)^2_0(a_0/a)^2$. This result was used in~\cite{BZDM} but
it does not incorporate the reheating effects, for which we also
need (\ref{eq:D3}). Then,
\begin{equation}
B=\tilde{B}_0\left(\frac{\sigma}{H}\right)_0
\left(\frac{\sigma}{H}\right)_{RH}
\left(\frac{\lambda_{\tilde{B}}}{\lambda_{H}}\right)^2_0
\left(\frac{a_0}{a}\right)^2\,,  \label{eq:Btot}
\end{equation}
where $RH$ marks the end of reheating. The difference in the above
is made by extra factor $(\sigma/H)_{RH}$. The latter is typically
very small and this accounts for the fact that the magnetic growth
rate during (late) reheating is considerably slower than that of the
radiation era. Therefore, not appreciating the role of (\ref{eq:D2})
and (\ref{eq:D3}) has also compromised the numerical results
of~\cite{BZDM}.

\section{On the gauge-invariance and the nonlinearity}
Studies of cosmological perturbations are known to suffer from the
gauge problem. The aim of~\cite{BZDM} is to provide a nonlinear
treatment of the gravito-magnetic interaction free from gauge
ambiguities. This meant integrating the magnetic induction equation
(i.e.~Eq.~(\ref{eq:dotB}) above) with respect to $B_a$ (see \S~IV
in~\cite{BZDM}). However, $B_a$ is treated as the perturbation of
$\tilde{B}_a$ (see \S~II.C in~\cite{BZDM}). The latter has nonzero
linear value and this makes $B_a$ a gauge-dependent vector at second
order by known theorems~\cite{SW}. In an attempt to circumvent the
problem the auxiliary variable $\beta_a$, with
$\beta_a\equiv\dot{B}_a+2\Theta B_a/3$ was temporarily introduced
(see \S~II.C in~\cite{BZDM}). This has zero linear value and is
therefore gauge-invariant at second order. Nevertheless, $\beta_a$
has been of little practical use because the authors still had to
solve for the gauge-dependent vector $B_a$ to extract any meaningful
information about the evolution of the field (see
Eqs.~(\ref{eq:D2}), (\ref{eq:D3}) and (\ref{eq:D6})
here).\footnote{If the authors had found a way of evaluating the
magnetic growth by means of $\beta_a$ exclusively, their results
would have been gauge-invariant. This was not possible however.
Moreover, at the ideal magnetohydrodynamics (MHD) limit $\beta_a$ is
all but redundant because $\beta_a\equiv \sigma_{ab}B^b$ (see
\S~II.C, IV.B in~\cite{BZDM} or Eq.~(\ref{eq:dotB}) here).} This
fact makes the analysis and the results of~\cite{BZDM} gauge
dependent.

The nonlinearity of~\cite{BZDM} is built on a set of four spacetimes
(see \S~II there). However, the new setting has not improved our
understanding of the interaction because still only the
gravito-magnetic effects of~\cite{TDM,T} are accounted for and all
other nonlinearities (including the magnetic effects on the shear)
are excluded. There is no new information in the nonlinear equations
of~\cite{BZDM}, relative to what is already encoded in the linear
formulae of~\cite{TDM,T}.\footnote{Expressions (4) and (10), the
latter with $\Lambda=0$, of~\cite{BZDM} are identical to Eqs.~(2b)
and (4) in~\cite{T}. Also, at the MHD limit, the magnetic wave
equation of~\cite{T} (see Eq.~(3) there) and expression (12)
in~\cite{BZDM} reduce to Eq.~(\ref{eq:dotB}) here. Moreover, the
constraints used in~\cite{BZDM} (see \S~II.B.2 there) are the
standard linear ones. These constraints ensure that the traceless
tensors remain transverse at all times, a highly nontrivial issue
for any truly nonlinear study.} This should not have happened and
the reason it does is the selective setting of~\cite{BZDM}, which
restricts the study and prevents it from going beyond the linear
level.

\section{Discussion}
The Maxwell-Weyl coupling and the possible energy transfer between
the two fields has a long research history. Recently this
interaction was proposed as a very efficient (resonant)
amplification mechanism of large-scale magnetic fields during the
poorly conductive stages of early reheating~\cite{TDM,T}. The same
mechanism was studied at the ideal MHD limit in~\cite{BZDM},
claiming that the gravito-magnetic interaction proceeds always as if
the universe were a perfect conductor when the background spatial
geometry is Euclidean. We explained here why the aforementioned
claim is not correct and how it restricts the generality
of~\cite{BZDM}. We also demonstrated that the aforementioned paper
did not properly monitor the gravito-magnetic interaction. In
particular, solutions which in~\cite{BZDM} were assigned to infinite
wavelengths only, were found to hold on all finite super-Hubble
scales. Also solutions that work only inside the horizon were
applied to all finite scales. As a result, the physical
interpretation and the numerical results of~\cite{BZDM} were
seriously compromised. With this comment we have attempted to
clarify these issues by correcting, where necessary, the mathematics
of the analysis and by explaining the physics of the interaction. In
the process we also considered and questioned the gauge-invariance
and the nonlinearity of the formalism proposed in~\cite{BZDM}.

\section*{Acknowledgments}
The author would like to thank Gerold Betschart, Caroline Zunckel,
Peter Dunsby and Mattias Marklund for their comments.

\end{document}